\documentclass[10pt,american,english,twocolumn,twoside,headsepline,footsepline]{scrartcl}

\usepackage{stmaryrd}
\usepackage{amsfonts}

\usepackage[T1]{fontenc}
\usepackage[utf8]{inputenc}
\usepackage{graphicx,times,amsmath,siunitx,prettyref} 
\usepackage{setspace}
\usepackage[disable]{todonotes}
\usepackage[numbers]{natbib}
\usepackage{abstract}

\newcommand{\mytitle}{Neuromorphic Learning towards Nano Second Precision}
\newcommand{\myrunningtitle}{Neuromorphic Learning}
\title{\mytitle}
\newcommand{\myauthor}{Pfeil et al.}
\newcommand{\kip}{\textsuperscript{1}}
\author{
Thomas Pfeil\footnotemark[1]~\footnotemark[2]~\kip
\and Anne-Christine Scherzer\kip
\and Johannes Schemmel\kip
\and Karlheinz Meier\kip
}

\usepackage[colorlinks=true,linkcolor=black,citecolor=black,urlcolor=black,filecolor=black,
pdffitwindow,pdftitle={\mytitle{}},pdfauthor={\myauthor},pdfsubject={},pdfkeywords={},bookmarks=false]{hyperref}

\usepackage{fixltx2e}
\usepackage[a4paper]{geometry}
\usepackage{dblfloatfix}

\usepackage{sectsty}
\allsectionsfont{\sf}

\usepackage{fancyhdr}
\makeatletter

\fancypagestyle{plain} 
{

\fancyhf{}
\fancyfoot[L]{}
\fancyfoot[C]{}
\fancyfoot[R]{}
}

\renewcommand\@seccntformat[1]{\csname the#1\endcsname.\quad}
\renewcommand*\section{\@startsection{section}{1}{\z@}%
{-1.7ex}
{0.7ex}
{\raggedsection\normalsize\bfseries\sffamily\color{blue}\nobreak\MakeUppercase}%
}
\renewcommand*\subsection{\@startsection{subsection}{2}{\z@}%
{\baselineskip}%
{0.25ex \@plus .2ex}%
{\raggedsection\small\bfseries\sffamily\nobreak\MakeUppercase}%
}
\renewcommand*\subsubsection{\@startsection{subsubsection}{3}{\z@}%
{\baselineskip}%
{0.25ex \@plus .2ex}%
{\raggedsection\small\bfseries\sffamily\itshape\nobreak}%
}
\setcapindent{0pt}
\setkomafont{captionlabel}{\bfseries\sffamily}
\setkomafont{caption}{\footnotesize\sffamily}
\renewcommand{\fnum@figure}{\textbf{FIGURE~\thefigure}}
\renewcommand{\fnum@table}{\textbf{TABLE~\thetable}}

\makeatother

\newcommand{\smalltodo}[2][]
{\todo[caption={#2}, #1]{\begin{spacing}{0.5}#2\end{spacing}}}

\newrefformat{cap}{\hyperref[#1]{Figure~\ref{#1}}}
\newrefformat{fig}{\hyperref[#1]{Figure~\ref{#1}}}
\newrefformat{sec}{\hyperref[#1]{Section~\ref{#1}}}
\newrefformat{sub}{\hyperref[#1]{Section~\ref{#1}}}
\newrefformat{tab}{\hyperref[#1]{Table \ref{#1}}}
\newrefformat{eq}{\hyperref[#1]{Equation~\ref{#1}}}
\newcommand{\mrmsub}[2]{#1_{\text{#2}}}
\global\long\def\taum{\mrmsub{\tau}{m}}
\global\long\def\taumhw{\taum^{\text{hw}}}
\global\long\def\delaysigma{\sigma}
\global\long\def\delayadd{\mrmsub{\Delta}{ITD}}
\global\long\def\weight{w}
\global\long\def\weighti{\weight_i}
\global\long\def\weightstart{\mrmsub{\weight}{start}}
\global\long\def\freq{f}
\global\long\def\period{T}
\global\long\def\prob{\mrmsub{p}{spike}}
\global\long\def\deltat{\Delta t}
\global\long\def\deltaw{\Delta w}
\global\long\def\delaydelta{\Delta d}
\global\long\def\delaydeltai{\delaydelta_i}
\global\long\def\vs{\nu}
\global\long\def\precision{\mrmsub{\sigma}{PL}}
\global\long\def\mem{\mrmsub{V}{m}}
\global\long\def\rate{r}
\global\long\def\gbase{g^{max}}
\global\long\def\gbasei{\gbase_i}
\global\long\def\gsyn{g}
\global\long\def\gsyni{\gsyn_i}
\global\long\def\aa{\mrmsub{a}{a}}
\global\long\def\ac{\mrmsub{a}{c}}
\global\long\def\ath{\mrmsub{a}{th}}
\global\long\def\phaseanglei{\Theta_i}

\begin{document}

\begin{titlepage}\thispagestyle{empty}\pdfbookmark[1]{Title}{TitlePage}

\maketitle

\begin{abstract}
\noindent Temporal coding is one approach to representing information in spiking neural networks.
An example of its application is the location of sounds by barn owls that requires especially precise temporal coding.
Dependent upon the azimuthal angle, the arrival times of sound signals are shifted between both ears.
In order to determine these interaural time differences, the phase difference of the signals is measured.
We implemented this biologically inspired network on a neuromorphic hardware system and demonstrate spike-timing dependent plasticity on an analog, highly accelerated hardware substrate.
Our neuromorphic implementation enables the resolution of time differences of less than \SI{50}{\nano\second}.
On-chip Hebbian learning mechanisms select inputs from a pool of neurons which code for the same sound frequency.
Hence, noise caused by different synaptic delays across these inputs is reduced.
Furthermore, learning compensates for variations on neuronal and synaptic parameters caused by device mismatch intrinsic to the neuromorphic substrate.
\end{abstract}

\vfill{}

\noindent$^{1}$ \parbox[t]{10cm}{Kirchhoff-Institute for Physics\\
Heidelberg University\\
Heidelberg, Germany}

\vspace{2em}

\noindent$^{*}$ Correspondence:\hspace{1em}\parbox[t]{10cm}{
Thomas Pfeil\\
Heidelberg University\\
Kirchhoff-Institute for Physics\\
Im Neuenheimer Feld 227\\
69120 Heidelberg, Germany\\
tel: +49-6221-549813\\
\href{mailto:thomas.pfeil@kip.uni-heidelberg.de}{thomas.pfeil@kip.uni-heidelberg.de}}

\vspace{2em}

\noindent$^{\dagger}$ Received funding by the European Union 7th Framework Programme under grant agreement no. 243914 (Brain-i-Nets).

\end{titlepage}

\section{Introduction}
\begin{figure}[t]
\centerline{\includegraphics[width=3.4in]{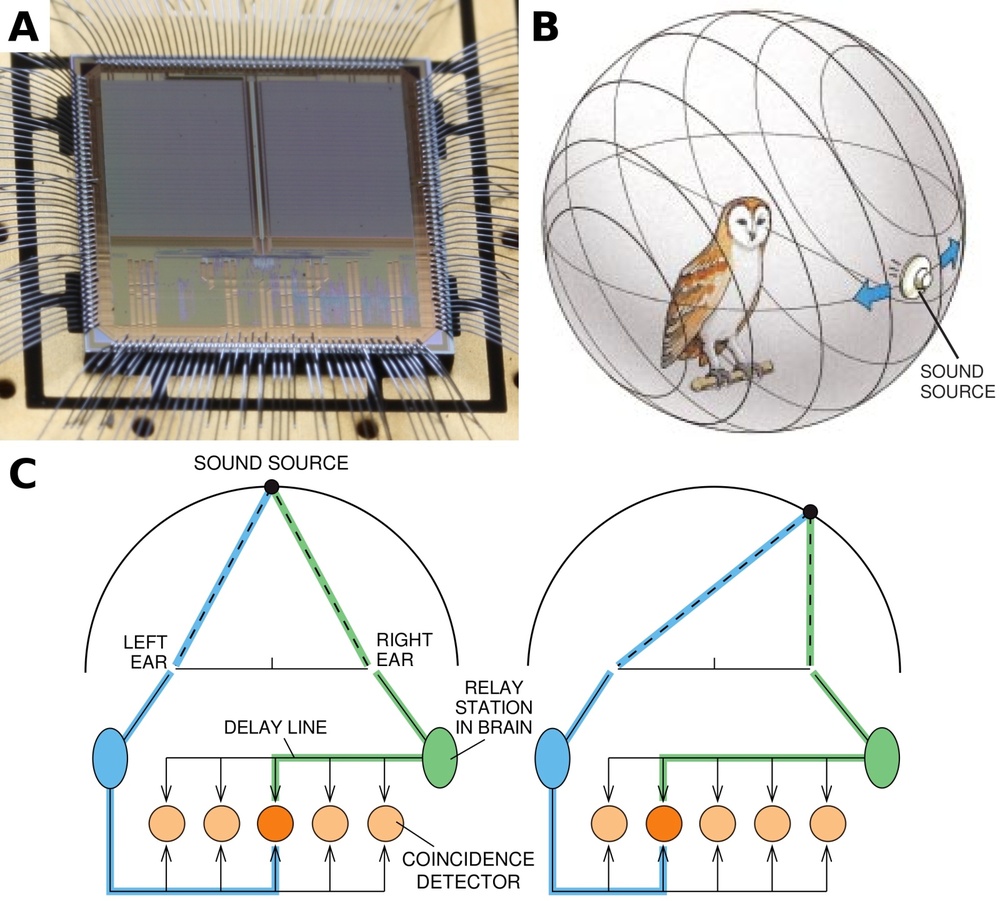}}
\caption{
\textbf{(A)}
Microphotograph of the neuromorphic chip (fabricated in a \SI{180}{\nano\meter} CMOS process with die size $\SI{5}{} \times \SI{5}{\milli\meter\squared}$).
\textbf{(B)}
A barn owl locates sound by measuring the interaural time difference of a presented sound signal.
\textbf{(C)}
Schematic of the auditory pathway of barn owls.
Neurons in the laminar nucleus detect coincidences between signals that arrive from both ears through fibres serving as delay lines (blue and green lines).
The firing rate of the neuron at which the signals arrive simultaneously has the highest firing rate (colored darkly).
Consequently, each neuron codes for an azimuthal sector.
Figure (B) and (C) are adapted from \cite{Konishi93_66}.
}
\label{fig:intro}
\end{figure}

Phase-locking has been shown to be one approach towards precise temporal coding in neural information processing \cite{Gerstner96} and is observed in the auditory pathway of barn owls \cite{Carr90_3227}.
Barn owls locate sounds by measuring the difference between the respective arrival times at both ears, the so-called interaural time difference (ITD), also known as the Jeffress model \cite{Jeffress48_35}.
A neuron in the laminar nucleus will fire at a high rate if it detects coincidences between the two periodic signals that code the same sound frequency at both ears (\prettyref{fig:intro}B and C).
In other words, the neuron will fire at a maximum rate if the signals from both ears arrive coherently.
This requires spike times that are ``locked'' to a specific phase of the sound signal.
The more precisely spikes are locked, the higher the temporal resolution for measuring ITDs is.
Previous studies have shown phase-locking with precision much smaller than the membrane time constants of the involved neurons \cite{Carr90_3227}. \smalltodo{check ref again}
However, synaptic delays differ across the neurons that code for the same sound frequency \cite{Carr90_3227}.
Coherence in the arrival time of signals can be restored by learning transmission delays \cite{Senn02_583} or by selecting synapses with simultaneous activity \cite{Gerstner96}.

In this study, we present an analog, neuromorphic implementation of a spiking neural network which selects inputs out of a broad distribution of transmission delays by means of an unsupervised, on-chip Hebbian learning rule.
The intrinsic, high acceleration factor of the neuromorphic substrate \cite{Pfeil13_11} allows for learning of phase-locking with \SI{100}{\nano\second} precision in hardware time domain.
Finally, the results of this on-chip synapse selection is applied to detect ITDs of less than \SI{50}{\nano\second}.

In addition to noise induced by variations in transmission delays, device mismatch in analog circuitries of neurons and synapses causes fixed-pattern noise on neural components.
This means, parameters vary between neurons and synapses, as, for example, in the membrane time constant and synaptic strength.

Both types of variations can be reduced by off-chip calibration routines.
In this study, they are compensated by on-chip learning mechanisms.
In contrast, an implementation without plasticity, but the same variations in neural components, can barely measure any time differences between two \SI{1}{\mega\hertz} signals in hardware time domain.

One of the first neuromorphic implementations for coincidence detection was a silicon replication of the auditory pathway \cite{Lazzaro89_47}.
Sound location is implemented by two synapses which transmit the signal from each ear.
However, this device does not support on-chip learning, and connectivity between neurons is hard-wired.
In this study, each ear is represented by a population of synapses rather than a single synapse.
Inspired by the biological example \cite{Carr90_3227} we add noise to the synaptic delays of this population.
We demonstrate that spike-timing dependent plasticity which is implemented in our neuromorphic hardware system de-noises the input.
The novelty of this system is to combine a dense integration of highly accelerated neurons with on-chip learning in each synapse \cite{Pfeil13_11} that allows for learning of coincidence detection with resolution higher than \SI{50}{\nano\second}.
Other approaches with re-configurable connectivity and on-chip plastic synapses have lower counts of neurons and synapses \cite{Indiveri06_211, Badoni06_4, Hafliger07_551}, mostly optimizied for low power consumption, or operate in biological real-time \cite{Seo11_1}.


\section{Network and hardware description}
\begin{figure}[t]
\centerline{\includegraphics[width=1.9in]{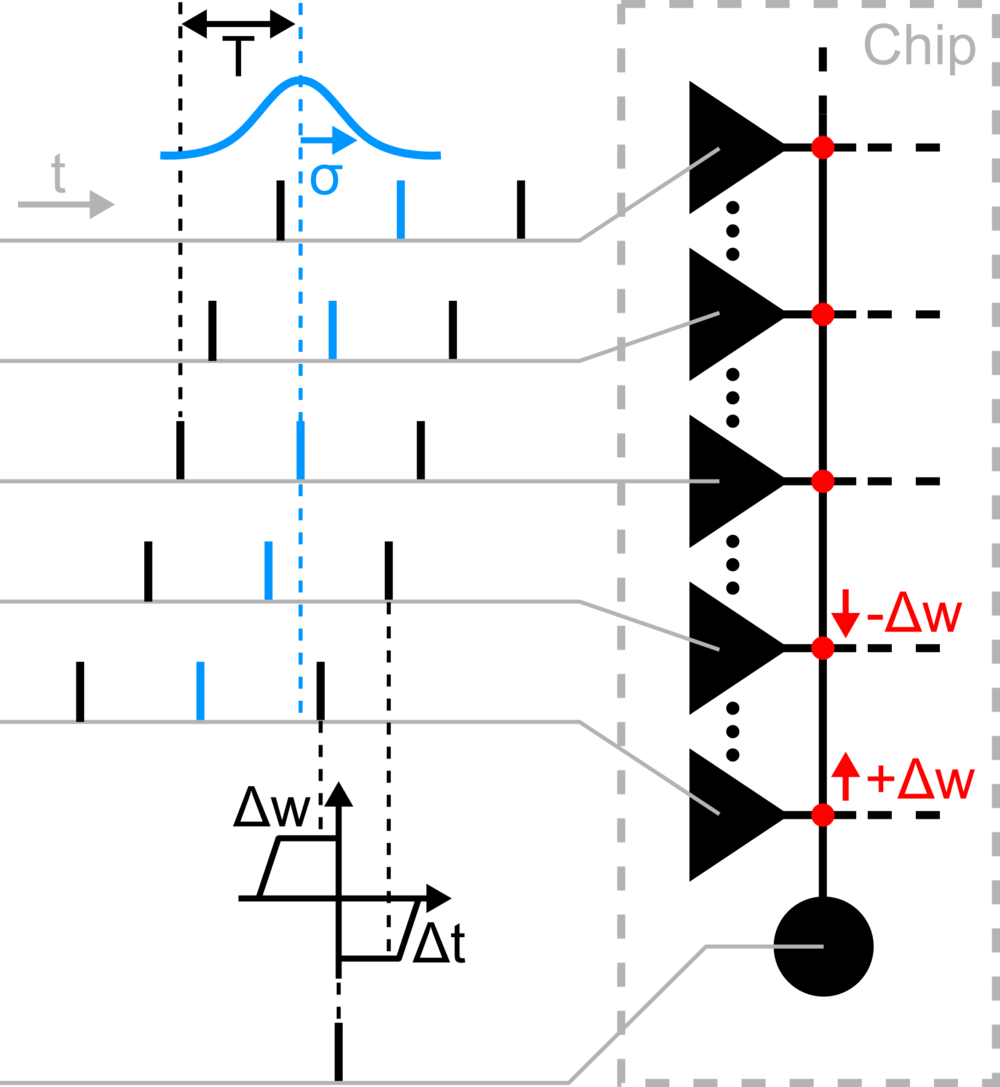}}
\caption{Network implementation on the neuromorphic chip.
The postsynaptic hardware neuron (black circle) receives input from $64$ presynaptic spike sources, whose spike times are generated on the host computer and transferred to the chip.
All inputs (black triangles) show regular spike times of identical frequency $\freq = \frac{1}{\period} = \SI{2}{\kilo\hertz}$, but each input is shifted in time following a Gaussian distribution with $\delaysigma = \SI{300}{\micro\second}$ (blue).
For simplicity we neglected $\prob < 1$ and jitter on spike times in this schematic.
When the postsynaptic neuron fires, each synapse (red circles) measures the correlations ($\deltat$) between its pre- and postsynaptic spikes.
During the next evaluation of these synapses, their weights $\weight$ are potentiated or depressed according to an additive rule ($\deltaw = \pm 1$, respectively).
}
\label{fig:sketch}
\end{figure}

The neuromorphic hardware system we used is designed as a re-configurable, universal neuromorphic substrate on which neural networks operate $10^4$ times faster than biological real-time (\prettyref{fig:intro}A, \cite{Pfeil13_11}).
It comprises pair-wise spike-timing dependent plasticity (STDP) in each of up to $256$ synapses per neuron \cite{Schemmel06_1}.
This type of Hebbian learning rule is adapted from measurements in biological tissue \cite{Markram97a, Bi98, Morrison08_459} and is described later in this section.

The network model implemented \emph{in silico} is inspired by the auditory pathway of barn owls \cite{Carr90_3227, Gerstner96}.
In this study, we investigate sound processing of a single frequency channel, exemplarily for any frequency of comparable scale.
A neuron in the nucleus laminaris receives bilateral input from several neurons in the cochlear nucleus magnocellularis (\prettyref{fig:sketch}).
The network selects those inputs, of which signals arrive coherently at the postsynaptic neuron, in order to improve the temporal coding of sound signals which is used for sound localization.

The delay of signal transmission $\delaydeltai$ for each connection with index $i$ is Gaussian distributed with standard deviation $\delaysigma$ \cite{Carr90_3227}.
Note that $\delaysigma \cdot 2$ is larger than the period $\period$ of the input signal and, consequently, neighboring volleys of spikes overlap (black and blue spikes in \prettyref{fig:sketch}).
The postsynaptic neuron is mimicked by a hardware neuron that approximates the conductance-based, leaky integrate-and-fire neuron model \cite{Pfeil13_11, Indiveri11_73}.
The presynaptic input is modeled by individual spike trains fed into the system.
\prettyref{fig:sketch} shows the allocation of hardware resources including the synaptic nodes, each of which comprises STDP.

As the membrane time constant of hardware neurons can not be configured to values below $\taumhw \approx \SI{200}{\nano\second}$ in hardware time domain, we assume an acceleration factor of $500$ throughout this study for better comparison with biological measurements.
Thus, all time constants, as well as experiment time, are given in biological time domain, if not stated otherwise.
For example, in the following, we use a neuron with $\taumhw = \SI{200}{\nano\second}$ in hardware time domain which translates to $\taum = \SI{100}{\micro\second}$ in biological time domain, as suggested in \cite{Gerstner96}.

\begin{figure}[t]
\centerline{\includegraphics[width=3.4in]{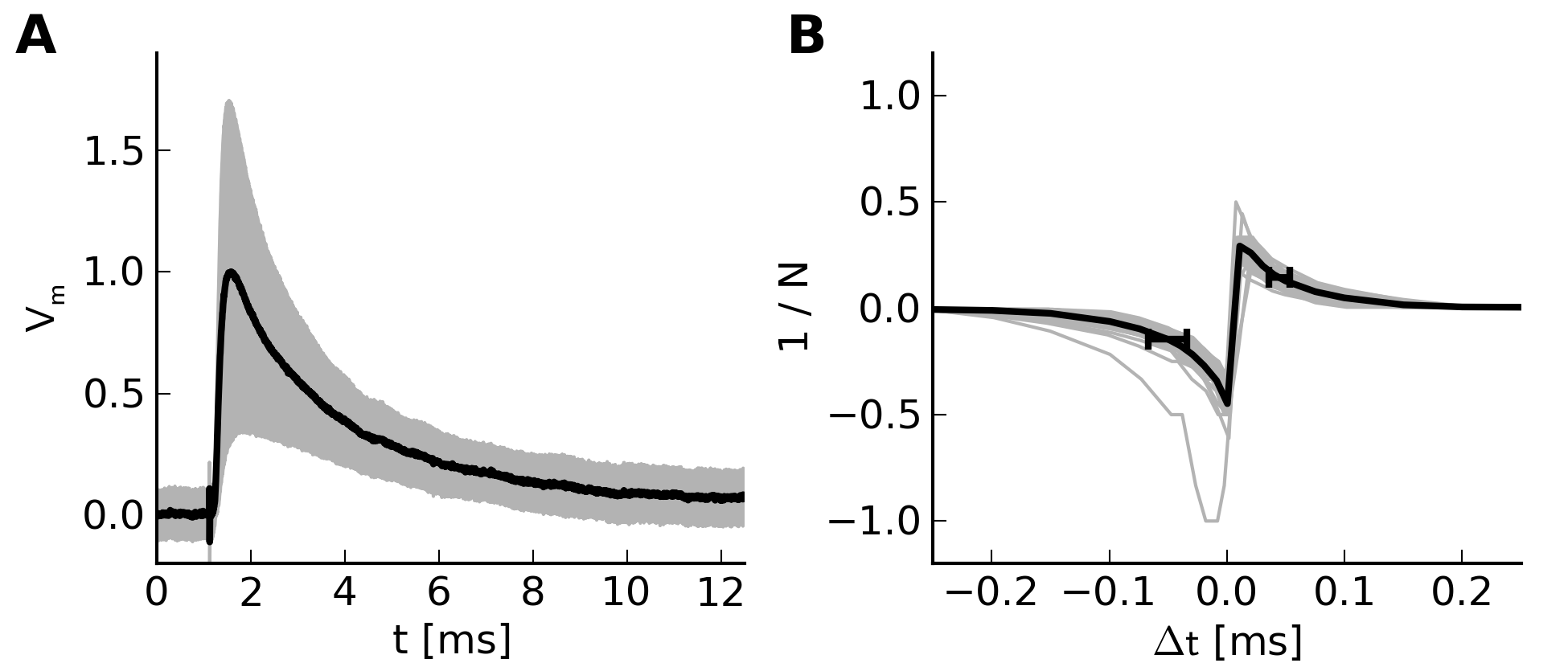}}
\caption{EPSPs and STDP learning windows of $64$ hardware synapses.
\textbf{(A)}
Each EPSPs is averaged over $100$ runs.
The mean and standard deviation over all synapses are depicted in black and gray, respectively.
The membrane potential $\mem$ is plotted in arbitrary units.
The area under these EPSPs has a ratio of standard deviation to mean of $\approx 50\%$.
\textbf{(B)}
Recording STDP learning windows (gray) is summarized as follows:
The inverse number $1 / N$ of spike pairs that need to be accumulated until a weight update is elicited is plotted against the time difference $\deltat$ between the pre- and postsynaptic spike (details are described elsewhere \cite{Pfeil12_90}).
A value of $N = 1$ means that one pre-post pair is already sufficient to mark this synapse to be updated.
The mean over all synapses and errors at half maximum ($0.044 \pm 0.009 \SI{}{\milli\second}$ and $-0.050 \pm 0.016 \SI{}{\milli\second}$) are depicted in black.
\smalltodo[inline]{EPSPs data is outdated}
}
\label{fig:hw_char}
\end{figure}

We stimulate the network with a pure tone similar to \cite{Gerstner96}.
For each presynaptic input spikes are drawn with probability $\prob = 0.35$ from a template of regular spike times with frequency $\freq = \SI{2}{\kilo\hertz}$ in biological time domain.
Additionally, each spike has a jitter following a Gaussian distribution with standard deviation \SI{40}{\micro\second}.
The individual transmission delay $\delaydeltai$ for each input with index $i$ is modeled by being added to all spike times of this input.

On hardware, the strength of a synaptic connection $\gsyni$ is the product of a conductance $\gbasei$ adjustable for each input stream (triangles in \prettyref{fig:sketch}) and a 4-bit digital weight $\weighti$ stored in each synapse:
\begin{equation}
\gsyni = \gbasei \weighti
\end{equation}
While $\gbasei$ is static throughout an emulation run, $\weighti$ is subject to STDP and can assume integer values between $0$ and $15$.
However, $\gbasei$ varies between synapses.
This is caused by device mismatch due to imperfections in the production process \cite{Pfeil13_11}.
Excitatory postsynaptic potentials (EPSPs) were recorded by measuring the impact of a single spike on the resting state of the postsynaptic membrane potential $\mem$ (\prettyref{fig:hw_char}A).
Their time course is configured to be as short as possible, and their amplitude ($\gbase$) is set to an intermediate value.
This results in a target firing rate of approximately \SI{1}{\kilo\hertz} in the final network implementation.

Synaptic weights $\weighti$ are modified by on-chip learning mechanisms at accelerated runtime described as follows:
The temporal correlations between spike pairs are measured and stored locally in each synapse by analog circuitry.
Correlations between pre-post and post-pre spike pairs are accumulated as charge on two capacitors, respectively.
In contrast to the local measurement and accumulation, the evaluation of these measurements is performed by controllers shared between synapses.
Thereby, the controller compares for each synapse the amount of charge on its capacitors as follows:
\begin{equation}
|\ac-\aa| > \ath
\label{eq:cond1}
\end{equation}
\begin{equation}
\ac - \aa > 0
\label{eq:cond2}
\end{equation}
where $\ac$ is the charge on the capacitor for pre-post, $\aa$ for post-pre spike pairs, and $\ath$ a configurable threshold.
If \prettyref{eq:cond1} is true the synaptic weight is updated and the capacitors are discharged.
Otherwise the weight stays unchanged and correlations are further accumulated without discharge of the capacitors.
If a weight update is elicited, the synaptic weight will be increased by one if \prettyref{eq:cond2} is true, otherwise it is decreased by one (in both cases with absorbing boundaries at the minimum and maximum value).
In fact, hardware STDP is not limited to this additive rule, but can be configured to any rule via look-up tables.
In the network presented, processing one synapse takes $\SI{1.5}{\micro\second}$ in hardware time domain, which has been shown to be sufficient for coincidence detection in small networks \cite{Pfeil12_90}.
For a detailed description of the implementation and configuration of hardware STDP see \cite{Schemmel06_1} and \cite{Pfeil12_90}.

Parameters for the learning window of STDP on hardware must be adjusted to meet two criteria:
On the one hand, the window should be broad in order to resolve a full period $\period$ of the input signal and to ensure a sufficiently high learning rate.
On the other hand, the number $N$ of spike pairs for close-by $\deltat$ should be distinguishable which is the case for overall large $N$.
In \prettyref{fig:hw_char}B a trade-off between both criteria is shown, because they can not be fulfilled together on the hardware.
Note that learning windows are subject to device mismatch too, due to their analog measurement and accumulation circuitry.
The width at half maximum is approximately \SI{0.05}{\milli\second} in biological time domain.
The time between successive weight updates for 64 inputs is $\SI{48}{\milli\second}$.

The experiment protocol is split into two steps.
First, the network is stimulated without any interaural time difference and on-chip plasticity is activated.
The network selects appropriate inputs by altering on-chip, synaptic weights.
Second, these learned synaptic weights are adopted for subsequent emulations, during which plasticity is turned off.
The ITD is varied and sound is located via the firing rate of the postsynaptic neuron.

In the first step, the spiking neural network shown in \prettyref{fig:sketch} is emulated for a duration of \SI{10}{\second}, while spike times of the postsynaptic neuron are recorded.
After emulation, the digital weights of all synapses are read out from the chip.

Network performance is measured by the vector strength $\vs$, which quantifies the precision of phase-locking at the postsynaptic site \cite{Goldberg69_613}.
Each spike time $t_i$ can be considered as a vector of unit length with a phase angle $\phaseanglei= 2 \pi \freq t_i$ as well as $x$ and $y$ components as follows:
\begin{equation}
(x_i, y_i) = (\sin(\phaseanglei), \cos(\phaseanglei))
\label{eq:vector}
\end{equation}
The vector strength is defined as the length of the mean vector across the overall number $N$ of postsynaptic spike times:
\begin{equation}
\vs = \frac{1}{N} \sqrt{
\left( \sum\nolimits_{i=1}^N x_i \right)^2 +
\left( \sum\nolimits_{i=1}^N y_i \right)^2 }
\end{equation}
It can assume values between $0$ and $1$.
It is minimal for randomly distributed spikes over time, and maximal for interspike intervals that are multiples of $\period$.
The angular dispersion of the mean vector translates to temporal precision \cite{Hill89_63}:
\begin{equation}
\precision = \frac{\sqrt{2 (1 - \vs)}}{2 \pi \freq}
\end{equation}

In the second step, the sensitivity of the postsynaptic neuron's firing rate to time differences between two input signals, e.g.\ ITDs, is determined as follows:
First, signal transmission delays and synaptic weights are adopted from the learning result of an emulation described above.
Second, afferent connections are randomly divided into two groups of equal size.
Third, time differences are modeled by adding an additional transmission delay to one of these groups.
Finally, this network is emulated with static synaptic weights, and the firing rate of the neuron is recorded.

\section{Hardware emulation results}
\begin{figure}[t]
\centerline{\includegraphics[width=3.4in]{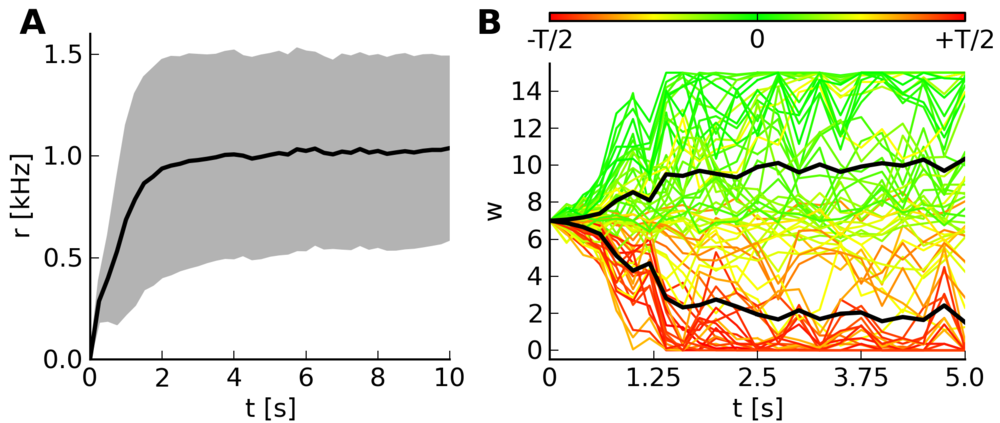}}
\caption{Learning process: Development of postsynaptic firing rates as well as synaptic weights over time.
\textbf{(A)}
The mean firing rate $\rate$ and standard deviation of $100$ emulations with \SI{10}{\second} duration in black and gray, repectively.
Transmission delays $\delaydelta$ and input spike times are re-drawn for each emulation.
\textbf{(B)}
Development of digital hardware weights $\weight$ over time for each synapse.
For technical convenience, each set of weights at time $t$ is recorded after an emulation with duration $t$.
For each time step weights are averaged over $20$ emulations, all with the same distribution of $\delaydelta$ and input spike times.
The color code indicates the difference to the average phase of the postsynaptic neuron.
}
\label{fig:devel}
\end{figure}

\begin{figure}[t]
\centerline{\includegraphics[width=3.4in]{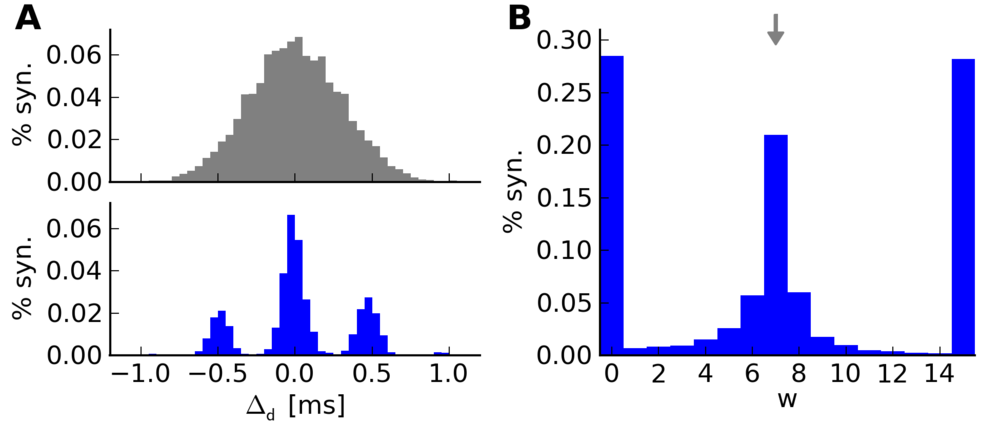}}
\caption{Selection of synapses and weights after on-chip learning.
\textbf{(A)}
Top: The distribution of signal transmission delays $\delaydelta$ before learning that is accumulated over the emulations shown in \prettyref{fig:devel}A.
The distribution is normalized such that its mean is $0$.
Bottom: The same distribution after learning, but only synapses with ${\weight > \weightstart}$ are shown.
The phase obtained by summing up the phase vectors (\prettyref{eq:vector}) of all spikes is subtracted from the transmission delays after each emulation.
In other words, transmission delays are normalized such that the postsynaptic neuron fires preferably at $\delaydelta = 0$.
\textbf{(B)}
The distribution of digital hardware weights $\weight$ after learning for the same emulations as in (A).
Before learning all weights were set to $\weightstart=7$ (arrow).
}
\label{fig:learned}
\end{figure}

Network emulations on hardware show precise, phase-locked spiking of the postsynaptic neuron.
At the beginning of an emulation, all synapses have the same weight, and the firing rate of the postsynaptic neuron is low.
The firing rate increases with increasing weights of those inputs that drive the neuron most (compare \prettyref{fig:devel}A to green traces in B).
Once strong synapses have evolved, phase-locking improves ($\vs$ increases, not shown) and synapses firing out of phase are weakend (red traces in \prettyref{fig:devel}B).
After approximately \SI{3}{\second}, the strong and weak synapses are in balance, and the postsynaptic firing rate saturates.
The variance of firing rates between emulations (\prettyref{fig:devel}A) is mostly caused by re-drawing the transmission delays for each emulation.
Variations on synapses bias the learning process.
For example, few strong synapses with improbable ($|\delaydelta| \gg 0$ in upper plot of \prettyref{fig:learned}A), but similar $\delaydelta$ may outperform many weak synapses with probable ($\delaydelta \approx 0$), but similar $\delaydelta$.
Nevertheless, learning compensates for these variations.

Synapses will be termed \emph{selected} if their weight after emulation exceeds their initial weight.
Synapses with similar transmission delays or delays which differ by multiples of $\period$ are active simultaneously, and preferably drive the postsynaptic neuron.
If a postsynaptic spike is elicited by these inputs, they are selected by the on-chip learning mechanism (compare \prettyref{fig:learned}A to \prettyref{fig:devel}B).
This periodical selection scheme with period $\period$ is due to the overlap of consecutive spike volleys (\prettyref{fig:sketch}).
However, not all synapses saturate to the minimum or maximum weight (\prettyref{fig:learned}B).
Some synaptic weights stay at their initial value.
This can be explained by the technical implementation of STDP on hardware.
If both correlations, pre-post and post-pre, accumulate at the same rate, no weight update will be triggered because the update controller evaluates the difference between both accumulations (see \prettyref{eq:cond1} and \cite{Schemmel06_1}).
\smalltodo[inline]{mention phase distribution after learning?}

\begin{figure}[t]
\centerline{\includegraphics[width=3.4in]{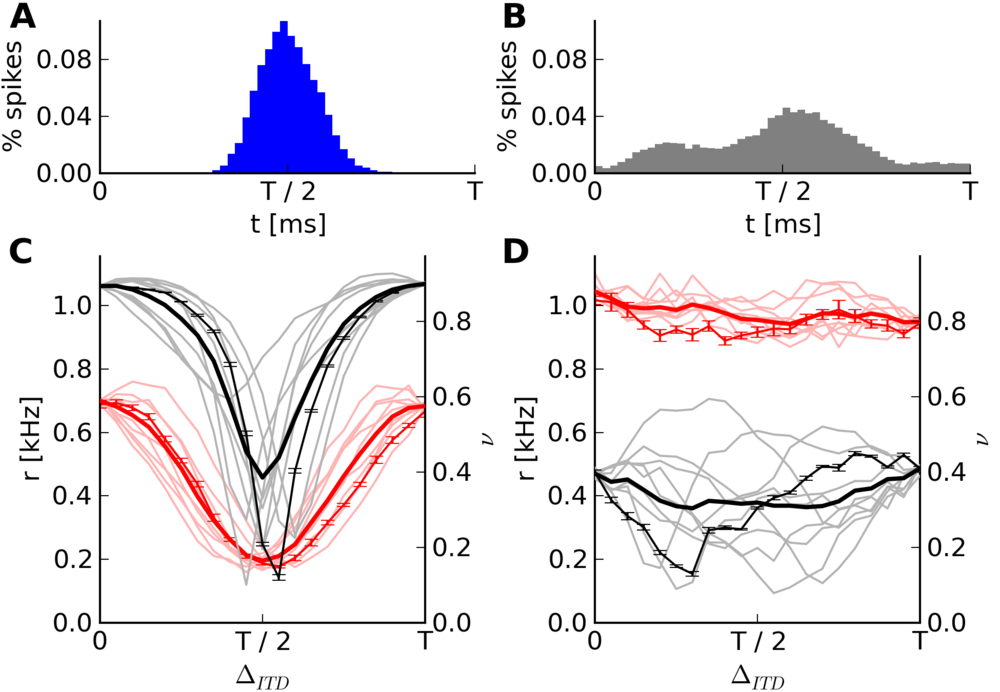}}
\caption{Phase-locking and the detection of interaural time differences.
\textbf{(A)}
Cyclic peristimulus time histograms (PSTHs) of postsynaptic spike times for one arbitrary emulation of those analyzed in \prettyref{fig:learned} ($\period = \SI{0.5}{\milli\second}$).
The vector strength is $\vs = 0.89$ ($\approx \SI{40}{\micro\second}$ precision), and the average vector strength over all $100$ emulations is $\bar{\vs} = 0.81 \pm 0.10$ ($\approx \SI{50}{\micro\second}$ precision).
\textbf{(B)}
PSTH for emulations of the same network and input as in (A), but STDP deactivated and $\weightstart$ adjusted to obtain similar postsynaptic firing rates ($\vs = 0.38$ and $\bar{\vs} = 0.43 \pm 0.19$).
\textbf{(C)}
Firing rate $\rate$ (bold red) and vector strength $\vs$ (bold black) averaged over $10$ emulations for one half of synapses receiving input delayed by $\delayadd$ compared to the other.
Single emulations are shown in light shades, each with another random division of the synapses, but the same spike times.
For one arbitrary random division the mean over $5$ emulations is shown in thin black lines.
Thereby, the standard deviation is the trial-to-trial variability.
The network has the same transmission delays and synaptic weights as in (A).
\textbf{(D)}
The same protocol as in (C), but with static synaptic weights adopted from (B).
}
\label{fig:itd}
\end{figure}

As an application, we show the detection of ITDs.
To this end, we split the resulting synapses of a single run of \prettyref{fig:learned} into two groups, one for the input from each ear.
Both, selected and unselected synapses are adopted.
The performance of phase-locking with $\delayadd = 0$ is shown in \prettyref{fig:itd}A.
Postsynaptic spikes occur preferably at the same phase with a vector strength of $\vs = 0.89$ which translates to a precision of $\precision \approx \SI{40}{\micro\second}$ in biological time domain.
Shifting the input of one group by $\delayadd$ deteriorates the precision of phase-locking and consequently reduces the postsynaptic firing rate (\prettyref{fig:itd}C).
At $\delayadd = \frac{\period}{2}$ the postsynaptic neuron receives alternating input from both groups, and the vector strength is on the same level as the control (compare \prettyref{fig:itd}C to D).
For the control we applied the same protocol as before (see \prettyref{fig:itd}A and C), but with a uniform weight distribution instead of previously learned synaptic weights.
Time differences of less than \SI{25}{\micro\second} can be resolved by the firing rate of the postsynaptic neuron (compare error bars of neighboring data points of thin red line in \prettyref{fig:itd}C).
In contrast, the firing rate of the control barely has a dependency on $\delayadd$ (thin red line in \prettyref{fig:itd}D).
This makes it difficult, if not impossible, to determine any time differences of \SI{2}{\kilo\hertz} signals.

\section{Conclusions}
\begin{figure}[t]
\centerline{\includegraphics[width=3.0in]{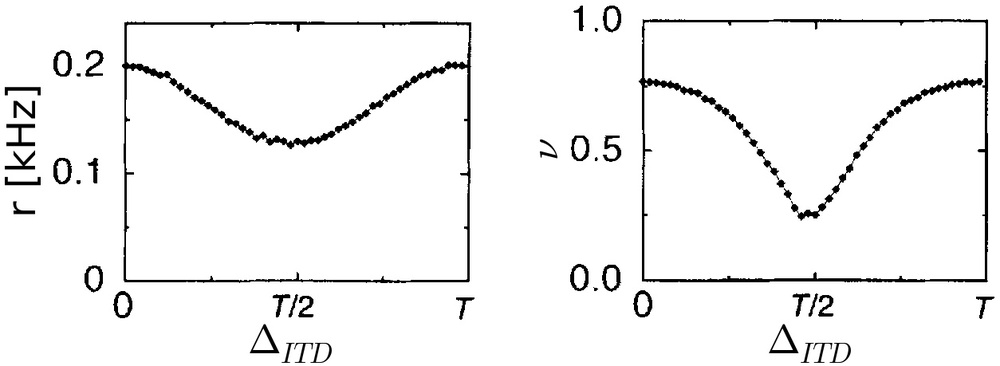}}
\caption{
Results of a similar network as described in this study, but simulated in software (adopted from \cite{Gerstner96}).
The firing rate $\rate$ and vector strength $\vs$ of the network is shown in dependence on the interaural time difference $\delayadd$.
The signal frequency is \SI{5}{\kilo\hertz}.
}
\label{fig:gerstner}
\end{figure}

We have presented an analog, neuromorphic network implementation that de-noises phase information and thereby learns to resolve time differences between two periodic stimuli of less than \SI{50}{\nano\second} in hardware time domain.
De-noising is realized by unsupervised, on-chip spike-timing dependent plasticity that improves coincidence detection and locks spike times to a specific phase of the input signal.
This results in precise phase information at the neuron site which enables the resolution of short phase differences, as for example those of sound signals from both ears.
The network performance is comparable to similar network models simulated in software (compare \prettyref{fig:itd}C to \prettyref{fig:gerstner}).
However, the absolute values for firing rate and vector strength differ due to different neuron, synapse, and STDP models, variations of model parameters on hardware, as well as a higher signal frequency in the reference publication \cite{Gerstner96}.

Additionally, learning does not only de-noise the input, but also compensates for variations between neural components (\prettyref{fig:hw_char}).
These variations are caused by device mismatch and are inherent in all neuromorphic systems with analog circuitry\smalltodo{cite?}.
Intrinsically weak synapses with simultaneous impact on the membrane potential can outperform intrinsically strong synapses.
This allows the measurement of short time differences, although the input is noisy and the neuromorphic substrate has variations in its neural components.
Performance may even be improved by a preceding off-chip calibration of synaptic strengths.
Furthermore, population coding reduces noise within signal transmissions by averaging across many unreliable components.

Although variations of neuronal components are measured in biology \cite{Marder06_563}, it is still unclear how robust a neural network has to be in order to perform computation on these components.
Large-scale neuromorphic systems, as described in \cite{Schemmel10_1947}, may particularly benefit from such self-adjusting, and hence robust, network implementations.
In further studies, the neuromorphic network could be embedded into robotic systems for processing sensory data of, for example, ultrasonic sound.
This would exploit the high acceleration factor of the neuromorphic system and its robust capability for handling noise and variations of neural components.

\smalltodo{add ``in silicon'' somewhere}

\bibliographystyle{neuralcomput_natbib}
\footnotesize
\bibliography{bib}
\normalsize

\end{document}